\documentclass[a4paper,12pt]{article}
\usepackage{amssymb,amsmath}
\usepackage{epsfig,fancyhdr,verbatim}

\usepackage{a4}
\usepackage{parskip}
\usepackage{appendix}
\usepackage{color,soul,rotating}
\usepackage{amsthm,amsxtra,amscd,relsize,multirow}
\usepackage[all,2cell]{xy}\UseAllTwocells

\newcommand{\sect}[1]{\setcounter{equation}{0}\section{#1}}

\def\g{\gamma}

\def\t{\tilde}

\def\cosh{\mathrm{cosh}}

\def\p{\partial}

\def\axs{AdS_5\times S^5}
\newcommand{\eq}[1]{\begin{equation} #1 \end{equation}}
\newcommand{\al}[1]{\begin{align} #1 \end{align}}

\begin{document}
%%%%%%%%%%%%%%%%%%%%
\begin{titlepage}
%\begin{flushright} \small TUW--11--11 \end{flushright}
\markright{\bf TUW--11--11}
%\vspace*{.5cm}
\title{Three-point correlators: Examples from Lunin-Maldacena background}

\author{D.~Arnaudov${}^{\star}$ and R.~C.~Rashkov${}^{\dagger,\star}$\thanks{e-mail:
rash@hep.itp.tuwien.ac.at.}
\ \\ \ \\
${}^{\star}$  Department of Physics, Sofia
University,\\
5 J. Bourchier Blvd, 1164 Sofia, Bulgaria
\ \\ \ \\
${}^{\dagger}$ Institute for Theoretical Physics, \\ Vienna
University of Technology,\\
Wiedner Hauptstr. 8-10, 1040 Vienna, Austria
}
\date{}
\end{titlepage}
%%%%%%%%%%%%%%%%%%%%%%%%%%%%%%%
%%%%%%%%%%%%%%%%%%%%%%%%%%%%%%%

\maketitle
\thispagestyle{fancy}

\begin{abstract}
Recently there has been progress on the calculation of three-point functions with two ``heavy'' operators via semiclassical methods. We extend this analysis to the case of the Lunin-Maldacena background, and examine the suggested procedure for several simple string solutions. By making use of AdS/CFT duality, we derive the relevant correlation functions of operators belonging to the dual $\,{\cal N}=1$ superconformal gauge theory, and recover an important relation connecting the dimensions of ``heavy'' states and the structure constants.
\end{abstract}

%%%%%%%%%%%%%%%%%%%%%%%%%%%%%%%%%
\sect{Introduction}
%%%%%%%%%%%%%%%%%%%%%%%%%%%%%%%%%

One of the most active fields of research in theoretical physics in recent years has been the correspondence between the large $N$ limit of gauge theories and string theory, and particularly the AdS/CFT correspondence~\cite{Maldacena}. Many impressive results from the duality between type IIB string theory on $\axs$ and ${\cal N}=4$ super Yang-Mills theory~\cite{Maldacena,GKP,Witten} have been obtained, but much more lies beyond our knowledge. One of the things that lack proper understanding is the calculation of three-point functions of operators with large quantum numbers at strong coupling ($\sqrt{\lambda}\gg1$). Although the problem remains unsolved in general, recently there has been progress in the semiclassical calculation of two-, three-, and four-point correlation functions with two ``heavy'' operators \cite{Janik:2010gc}--\cite{Bozhilov:2011}. Motivated by these studies, we consider the correlation functions of two ``heavy'' operators and one dilaton in $\gamma$-deformed SYM, applying the methods suggested in \cite{Janik:2010gc,Costa:2010}. The dual ${\cal N}=1$ field theory arises as an exactly marginal deformation of ${\cal N}=4$ SYM \cite{Leigh:1995,Berenstein:2000}.

The paper is organized as follows. In the next section we give a short review of the procedure for computing semiclassically three-point functions with dilaton ``light'' operator. Next, we proceed with the calculation of correlators corresponding to some simple string solutions in Lunin-Maldacena background. We conclude with a brief discussion on the results.

%%%%%%%%%%%%%%%%%%%%%%%%%%%%%%%%%%%%%%%%%%%%%%
\sect{Calculation of three-point correlators}
%%%%%%%%%%%%%%%%%%%%%%%%%%%%%%%%%%%%%%%%%%%%%%

Let us consider backgrounds of the form $AdS_5\times Y$, where $Y$ is a compact five-dimensional manifold. We will use Poincare coordinates $(z,x)$ for the AdS part, so the boundary will be a four-dimensional Minkowski space with coordinates $x$. As was shown in \cite{Costa:2010}, the correlation function of two ``heavy'' boundary operators and the dilaton $\phi$ at strong coupling assumes the following form
\eq{
\langle{\cal O}_A(x_i){\cal O}^*_A(x_f){\cal L}(y)\rangle\approx\frac{I_\phi[\bar{X}(\tau,\sigma),\bar{s};y]}{|x_i-x_f|^{2\Delta_A}}\,,
}
where we have used that the dilaton field sources the Lagrangian density near the boundary, and $\Delta_A$ is the scaling dimension of the ``heavy'' operators. With a slight abuse of notation
\eq{
I_\phi[\bar{X},\bar{s};y]=i\int_{-\bar{s}/2}^{\bar{s}/2}d\tau\int d\sigma\left.
\frac{\delta S_P[\bar{X},\bar{s},\phi]}{\delta\phi}\right|_{\phi=0}\,K_\phi(\bar{X};y)\,.
\label{Iphi}
}
This equation can be obtained by taking a functional derivative of the partition function, and then setting the dilaton to zero. The bulk-to-boundary propagator in \eqref{Iphi} comes from the supergravity part of the partition function, and has the following form \cite{Freedman:1998}
\eq{
K_\phi(\bar{X};y)=K_{\phi}(z(\tau),x(\tau);y)=\frac{6}{\pi^2}\!\left(\frac{z(\tau)}{z^2(\tau)-(x(\tau)-y)^2}\right)^4\!\!.
}
In addition, $\bar{s}$ is the saddle-point value of the modular parameter $s$ on the worldsheet cylinder, whose minimization of area gives $\langle{\cal O}_A(x_i){\cal O}^*_A(x_f)\rangle$. $\bar{X}$ stands for a classical solution to the equations of motion, which corresponds to an operator with large quantum numbers in the dual gauge theory, namely ${\cal O}_A$. In what follows, we will confine ourselves to solutions which are point-particle in AdS. They are described with
\begin{subequations}
\label{xzparticle}
\begin{align}
z&=z(\tau)=R/\cosh\,\kappa\tau\,,\\
x&=x(\tau)=R\tanh\kappa\tau+x_0\,.
\end{align}
\end{subequations}
As was shown in \cite{Janik:2010gc}
\eq{
\label{kappabc}
\kappa\approx\frac{2}{s}\ln\frac{x_f}{\varepsilon}\,,\qquad R\approx x_0\approx\frac{x_f}{2}\,,
}
where $\varepsilon$ is an ultraviolet cutoff, and $\kappa$ is defined through $t=\kappa\tau$. Also, the Polyakov action in conformal gauge is
\eq{
S_P[\bar{X},\bar{s},\phi]=-g\int_{-\bar{s}/2}^{\bar{s}/2}d\tau\int d\sigma\,e^{\phi/2}
\left(\eta^{\alpha\beta}\partial_\alpha\bar{X}^A\partial_\beta\bar{X}^Bg_{AB}-\epsilon^{\alpha\beta}\partial_\alpha\bar{X}^A\partial_\beta\bar{X}^BB_{AB}\right),
\label{Polyakov}
}
where $g\equiv\frac{\sqrt{\lambda}}{4\pi}$ is the coupling, $g_{AB}$ is the background metric, $\epsilon^{01}=1$, and $B_{AB}$ is the NS-NS 2-form field.

The detailed analysis in \cite{Janik:2010gc} shows that there is a subtlety in obtaining the string propagator, so that the classical solution for the cylinder coincides with the classical state. Therefore, when calculating $\bar{s}$ we have to consider
\begin{equation}
\tilde{S}_P=S_P-\int_{-s/2}^{s/2}d\tau\int d\sigma\,\Pi^A\dot{X}_A\,,
\label{NewAction}
\end{equation}
which is equal to minus the integral of the Hamiltonian. It was shown in \cite{Bak:2011} that, strictly speaking, we have to use the Routhian instead of minus the Hamiltonian, but they coincide in our case.

Also, the following relation is expected to hold (analyzed for $\axs$ in \cite{Costa:2010,Roiban:2010})
\eq{
\langle{\cal O}_A(0){\cal O}^*_A(x_f){\cal L}(y)\rangle\approx-\frac{1}{2\pi^2}\frac{g^2\partial\Delta_A}{\partial g^2}\frac{x_f^{4-2\Delta_A}}{y^4(x_f-y)^4}\,.
\label{aLAA}
}

%%%%%%%%%%%%%%%%%%%%%%%%%%%%%%%%%%%%%%%%%%%%%%%%%%%%%%%%%%%%%
\sect{Three-point functions from Lunin-Maldacena}
%%%%%%%%%%%%%%%%%%%%%%%%%%%%%%%%%%%%%%%%%%%%%%%%%%%%%%%%%%%%%

In this section we apply the methods described in the previous one to particular string solutions in Lunin-Maldacena background. The metric and $B$-field are given by \cite{Lunin:2005}
\al{\label{metric}
ds^2_{\rm str}/R^2_{\rm str}
&=\frac{dz^2+dx^2}{z^2}+\sum^3_{i=1}(d\rho_i^2+G\rho_i^2d\phi_i^2)+\t\g^2G\rho_1^2\rho_2^2\rho_3^2\left(\sum^3_{i=1}d\phi_i\right)^2\!\!,\\
\label{bfield}
B_2/R^2_{\rm str}&=\t\g G(\rho_1^2\rho_2^2d\phi_1d\phi_2+\rho_2^2\rho_3^2d\phi_2d\phi_3+\rho_3^2\rho_1^2d\phi_3d\phi_1)\,,\\ \nonumber\\
G^{-1}&\equiv1+\t\g^2(\rho_1^2\rho_2^2+\rho_2^2\rho_3^2+\rho_1^2\rho_3^2)\,,\quad\rho_1=\sin\alpha\cos\theta\,,\ \rho_2=\sin\alpha\sin\theta\,,\ \rho_3=\cos\alpha\,.
\nonumber
}
Here $\t\g=\g\sqrt{\lambda}$ is the deformation parameter, and we shall assume that it is independent of the coupling.

%%%%%%%%%%%%%%%%%%%%%%%%%%%%%%%%%%%%%%%%%%%%%%%%%%%%%%%%%%%%%%%%%%%%%%%%%%
\subsection{Circular rotating string with two equal spins in $S^5_{\t\g}$}
%%%%%%%%%%%%%%%%%%%%%%%%%%%%%%%%%%%%%%%%%%%%%%%%%%%%%%%%%%%%%%%%%%%%%%%%%%

The simplest nontrivial spinning string solution in Lunin-Maldacena background is the circular rotating string with two equal spins \cite{Frolov:2005}. It is contained in the subspace
\eq{\label{sol1}
\alpha=\frac{\pi}{2}\,,\quad\theta=\frac{\pi}{4}\,,\quad\phi_1=\omega\tau+m\sigma\,,\quad\phi_2=\omega\tau-m\sigma\,.
}
Assuming that both angular momenta $J_{1,2}$ corresponding to the angles $\phi_{1,2}$ are positive, it follows that
\eq{
J_1=J_2\equiv\frac{J}{2}\,,\quad J=4\pi gG\!\left(\omega-\frac{\t\g m}{2}\right),\quad G^{-1}=1+\frac{\t\g^2}{4}\,.
}
This solution has a smooth limit to the undeformed case $\t\g\rightarrow0$. The corresponding energy is
\eq{
E=\sqrt{J^2+\!\left(4\pi gm+\frac{\t\g J}{2}\right)^2}\,.
}
The ``heavy'' operators in the dual gauge theory are of the kind ${\cal O}_A\sim{\rm Tr}(\Phi_1^J\Phi_2^J)$, where $\Phi_i$ are complex scalars corresponding to the bulk quantities $\Phi_i=\rho_ie^{i\phi_i}$.

Let us apply now the procedure outlined briefly in the previous section. For the Polyakov action in this case we have
\begin{align}
S_P[\bar{X},s,\phi=0]&=\frac{g}{2}\int_{-s/2}^{s/2}d\tau\int_{0}^{2\pi}d\sigma\left[2\kappa^2+G(\dot{\phi_1}^2-{\phi_1^\prime}^2+\dot{\phi_2}^2-{\phi_2^\prime}^2)
+\t\g G(\dot{\phi_1}\phi^{\prime}_2-\phi^{\prime}_1\dot{\phi_2})\right]\nonumber\\
&=2\pi g\!\left[\frac{4}{s^2}\log^2\frac{x_f}{\varepsilon}+G(\omega^2-m^2-\t\g\omega m)\right]\!s\,,
\end{align}
where relation \eqref{kappabc} is used. The modified action \eqref{NewAction} in this case takes the form
\eq{
\tilde{S}_P[\bar{X},s,\phi=0]=2\pi g\!\left[\frac{4}{s^2}\log^2\frac{x_f}{\varepsilon}-G(\omega^2+m^2)\right]\!s\,.
}
The saddle point with respect to the modular parameter $s$ is given by
\eq{
\label{saddlecircularI}
\bar{s}=\frac{2i}{\sqrt{G(\omega^2+m^2)}}\log\frac{\varepsilon}{x_f}\,,
}
which together with \eqref{kappabc} implies the Virasoro constraint $\kappa=i\sqrt{G(\omega^2+m^2)}$. Therefore, it follows that
\eq{
\langle{\cal O}_A(0){\cal O}^*_A(x_f)\rangle\sim e^{i\tilde{S}_P[\bar{X},\bar{s},\phi=0]}=\left(\frac{\varepsilon}{x_f}\right)^{8\pi g\sqrt{G(\omega^2+m^2)}}.
}
It can be shown straightforwardly that $4\pi g\sqrt{G(\omega^2+m^2)}=E$, so that $\Delta_A=E$ in agreement with the AdS/CFT prediction.

Now we turn to the derivation of the three-point function. First we evaluate
\begin{align}
I_\phi[\bar{X},s;y]&=\frac{3ig}{\pi^2}\int_{-s/2}^{s/2}d\tau\!\int_{0}^{2\pi}\!d\sigma\left[\kappa^2
+G(\omega^2-m^2-\t\g\omega m)\right]\!\left(\frac{z}{z^2+(x-y)^2}\right)^4\nonumber\\
&=\frac{igs}{4\pi\log\frac{x_f}{\varepsilon}}\!\left[\frac{4}{s^2}\log^2\frac{x_f}{\varepsilon}+G(\omega^2-m^2-\t\g\omega m)\right]\!\frac{x_f^4}{y^4(x_f-y)^4}\,.
\end{align}
At the saddle point \eqref{saddlecircularI} we get
\eq{
I_\phi[\bar{X},\bar{s};y]=-\frac{gm}{\pi E}\!\left(4\pi gm+\frac{\t\g J}{2}\right)\!\frac{x_f^4}{y^4(x_f-y)^4}\,.
}
Therefore, the final expression for the three-point correlation function is
\eq{
\langle{\cal O}_A(0){\cal O}^*_A(x_f){\cal L}(y)\rangle\approx-\frac{gm\sqrt{E^2-J^2}}{\pi E}\frac{x_f^{4-2E}}{y^4(x_f-y)^4}\,.
\label{3pointcirI}
}
With some effort it can be seen that indeed \eqref{aLAA} holds, provided that $J$ and $m$ are kept constant.

%%%%%%%%%%%%%%%%%%%%%%%%%%%%%%%%%%%%%%%%%%%%%%%%%%%%%%%%%%%%%%%%%%%%%%%%%%
\subsection{Giant magnon and spiky string}
%%%%%%%%%%%%%%%%%%%%%%%%%%%%%%%%%%%%%%%%%%%%%%%%%%%%%%%%%%%%%%%%%%%%%%%%%%

The simplest giant magnon (spiky string) solution in Lunin-Maldacena background assumes the following form \cite{Bobev:2007}
\eq{\label{sol3}
\alpha=\frac{\pi}{2}\,,\quad\theta=\theta(\eta)\,,\quad\phi_1=\omega_1\tau+f_1(\eta)\,,\quad\phi_2=\omega_2\tau+f_2(\eta)\,,
}
where $\eta=\alpha\sigma+\beta\tau$. We require that one of the turning points of $\theta$ occurs at $\theta=\pi/2$. Then two cases can be differentiated -- giant magnon and spiky string.

\paragraph{Giant magnon}
For the case of giant magnon we have the following equations of motion
\begin{subequations}
\al{
f_1'&=-\frac{\cos^2\theta}{\alpha^2-\beta^2}\!\left(\frac{\beta\omega_1}{\sin^2\theta}+\t\g\alpha\omega_2+\t\g^2\beta\omega_1\right),\\
f_2'&=\frac{\beta\omega_2+\t\g\alpha\omega_1\sin^2\theta}{\alpha^2-\beta^2}\,,\\
\theta'&=\frac{\alpha\Omega_0}{(\alpha^2-\beta^2)}\frac{\cos\theta}{\sin\theta}\sqrt{\sin^2\theta-\sin^2\theta_0}\,,
}
\end{subequations}
where we have defined
\eq{
\sin\theta_0=\frac{\beta\omega_1}{\alpha\Omega_0}\,,\qquad\qquad\Omega_0=\sqrt{\omega_1^2-\left(\omega_2+\t\g\frac{\beta\omega_1}{\alpha}\right)^2}.
}
The corresponding dispersion relation is
\eq{
E-J_1=\sqrt{J_2^2+16g^2\sin^2\!\left(\frac{\Delta\phi_1}{2}-\t\g\frac{J_2}{4g}\right)}\,,
}
where $\Delta\phi_1$ is the angular difference, which is related to the momentum $p$ of the magnon excitation on the spin chain in the following way
\eq{
\Delta\phi_1=p-2\pi\t\g+\t\g\frac{J_2}{2g}\,.
}

Let us apply now the procedure outlined briefly in the previous section. For the Polyakov action we have
\begin{align}\nonumber
&S_P[\bar{X},s,\phi=0]=\frac{gs}{\alpha}\int_{-\infty}^{\infty}d\eta\,\bigg\{\frac{4}{s^2}\log^2\frac{x_f}{\varepsilon}
+(\beta^2-\alpha^2)\theta^{\prime2}+G\sin^2\theta\big[(\omega_1+\beta f_1')^2-\alpha^2f_1^{\prime2}\big]\\
&+G\cos^2\theta\big[(\omega_2+\beta f_2')^2-\alpha^2f_2^{\prime2}\big]+2\t\g G\alpha\sin^2\theta\cos^2\theta(\omega_1f_2'-\omega_2f_1')\bigg\}\,,
\label{polmagnon}
\end{align}
where $G^{-1}=1+\t\g^2\sin^2\theta\cos^2\theta$, and again relation \eqref{kappabc} is used. The modified action \eqref{NewAction} in this case is
\al{\nonumber
&\tilde{S}_P[\bar{X},s,\phi=0]=\frac{gs}{\alpha}\int_{-\infty}^{\infty}d\eta\,\bigg\{\frac{4}{s^2}\log^2\frac{x_f}{\varepsilon}
-(\alpha^2+\beta^2)\theta^{\prime2}-G\sin^2\theta\big[\alpha^2f_1^{\prime2}+(\omega_1+\beta f_1')^2\big]\\
&-G\cos^2\theta\big[\alpha^2f_2^{\prime2}+(\omega_2+\beta f_2')^2\big]\bigg\}
=\frac{gs}{\alpha}\int_{-\infty}^{\infty}d\eta\left(\frac{4}{s^2}\log^2\frac{x_f}{\varepsilon}-\omega_1^2\right).
}
The saddle point with respect to the modular parameter $s$ is given by
\eq{
\label{saddlemagnon}
\bar{s}=\frac{2i}{\omega_1}\log\frac{\varepsilon}{x_f}\,,
}
which together with \eqref{kappabc} implies $\kappa=i\omega_1$. Therefore, in agreement with the AdS/CFT prediction
\eq{
\langle{\cal O}_A(0){\cal O}^*_A(x_f)\rangle\sim\left(\frac{\varepsilon}{x_f}\right)^{2E}\ \
\Longrightarrow\ \ \Delta_A=\frac{2g\omega_1}{\alpha}\int_{-\infty}^{\infty}d\eta=\frac{4g\omega_1}{\alpha}\int_{\theta_0}^{\pi/2}\frac{d\theta}{\theta'}=E\,.
}

Now we turn to the derivation of the three-point function. We evaluate at the saddle point \eqref{saddlemagnon}
\begin{align}
&I_\phi[\bar{X},\bar{s};y]=-\frac{g}{\pi^2\omega_1}\frac{x_f^4}{y^4(x_f-y)^4}\int_{\theta_0}^{\pi/2}\frac{d\theta}{\theta'}\Big[
\alpha\theta^{\prime2}+G\alpha\sin^2\theta f_1^{\prime2}+G\alpha\cos^2\theta f_2^{\prime2}\\
&-\t\g G\sin^2\theta\cos^2\theta(\omega_1f_2'-\omega_2f_1')\Big]
=-\frac{1}{4\pi^2}\!\left(E-J_1-\frac{J_2^2}{E-J_1}+\t\g\frac{\beta J_2}{\alpha}\right)\!\frac{x_f^4}{y^4(x_f-y)^4}\,.
\nonumber
\end{align}
Thus, the final expression for the three-point correlator is
\eq{
\langle{\cal O}_A(0){\cal O}^*_A(x_f){\cal L}(y)\rangle\approx-\frac{1}{4\pi^2}\!\left(E-J_1-\frac{J_2^2}{E-J_1}+\t\g\frac{\beta J_2}{\alpha}\right)\!\frac{x_f^{4-2E}}{y^4(x_f-y)^4}\,.
\label{3pointmagnon}
}
It can be seen straightforwardly that \eqref{aLAA} holds, provided that $J_1,J_2,\Delta\phi_1$ are constant.

\paragraph{Spiky string}
For the case of spiky string the equations of motion are
\begin{subequations}
\al{
f_1'&=\frac{1}{\alpha^2-\beta^2}\!\left(\beta\omega_1-\frac{\alpha^2\omega_1}{\beta\sin^2\theta}-\t\g\alpha\sqrt{\omega_1^2-\Omega_1^2}\cos^2\theta\right),\\
f_2'&=\frac{\beta\omega_2+\t\g\alpha\omega_1\sin^2\theta}{\alpha^2-\beta^2}\,,\\
\theta'&=\frac{\alpha\Omega_1}{(\alpha^2-\beta^2)}\frac{\cos\theta}{\sin\theta}\sqrt{\sin^2\theta-\sin^2\theta_1}\,,
}
\label{eqspiky}
\end{subequations}
where
\eq{
\sin\theta_1=\frac{\alpha\omega_1}{\beta\Omega_1}\,,\qquad\qquad\Omega_1=\sqrt{\omega_1^2-\left(\omega_2+\t\g\frac{\alpha\omega_1}{\beta}\right)^2}.
}
The energy in this case is
\eq{
E=4g\arcsin\frac{\sqrt{J_1^2-J_2^2}}{4g}+2g\Delta\phi_1+\t\g J_2\,,
}
where again $\Delta\phi_1$ is the angular difference.

For the Polyakov action we obtain again \eqref{polmagnon} with $G^{-1}=1+\t\g^2\sin^2\theta\cos^2\theta$, where equations \eqref{eqspiky} should be plugged in. The modified action \eqref{NewAction} takes the form
\eq{
\tilde{S}_P[\bar{X},s,\phi=0]=\frac{gs}{\alpha}\int_{-\infty}^{\infty}d\eta\left(\frac{4}{s^2}\log^2\frac{x_f}{\varepsilon}-\frac{\alpha^2\omega_1^2}{\beta^2}\right).
}
The saddle point with respect to the modular parameter $s$ is provided by
\eq{
\label{saddlespiky}
\bar{s}=\frac{2i\beta}{\alpha\omega_1}\log\frac{\varepsilon}{x_f}\,,
}
which implies $\kappa=i\alpha\omega_1/\beta$. Thus, one can obtain
\eq{
\langle{\cal O}_A(0){\cal O}^*_A(x_f)\rangle\sim\left(\frac{\varepsilon}{x_f}\right)^{2E}\ \
\Longrightarrow\ \ \Delta_A=\frac{2g\omega_1}{\beta}\int_{-\infty}^{\infty}d\eta=\frac{4g\omega_1}{\beta}\int_{\pi/2}^{\theta_1}\frac{d\theta}{\theta'}=E\,,
}
which again is in agreement with the AdS/CFT prediction.

In order to derive the three-point correlator we evaluate at the saddle point \eqref{saddlespiky}
\begin{align}
&I_\phi[\bar{X},\bar{s};y]=-\frac{\beta g}{\pi^2\alpha\omega_1}\frac{x_f^4}{y^4(x_f-y)^4}\int_{\pi/2}^{\theta_1}\frac{d\theta}{\theta'}\Big[
\alpha\theta^{\prime2}+G\alpha\sin^2\theta f_1^{\prime2}+G\alpha\cos^2\theta f_2^{\prime2}\\
&-\t\g G\sin^2\theta\cos^2\theta(\omega_1f_2'-\omega_2f_1')\Big]
=-\frac{1}{4\pi^2}\bigg[E+\frac{\beta J_1}{\alpha}\!\left(1-\frac{J_2^2}{J_1^2}\right)-\t\g J_2\bigg]\frac{x_f^4}{y^4(x_f-y)^4}\,.
\nonumber
\end{align}
Therefore, we obtain finally for the three-point function
\eq{\label{3pointspiky}
\langle{\cal O}_A(0){\cal O}^*_A(x_f){\cal L}(y)\rangle\approx-\frac{1}{4\pi^2}\bigg[E+\frac{\beta J_1}{\alpha}\!\left(1-\frac{J_2^2}{J_1^2}\right)-\t\g J_2\bigg]\frac{x_f^{4-2E}}{y^4(x_f-y)^4}\,.
}
Again it can be seen that \eqref{aLAA} holds, provided that $J_1,J_2,\Delta\phi_1$ are kept independent of the coupling.

%%%%%%%%%%%%%%%%%%%%%%%%%%%%%%%%%%%%%%%%%%%
\sect{Conclusion}
%%%%%%%%%%%%%%%%%%%%%%%%%%%%%%%%%%%%%%%%%%%

In this paper we apply the ideas for the calculation of correlation functions developed for $\axs$ in \cite{Janik:2010gc,Costa:2010} to the case of Lunin-Maldacena background. We examine the method in the cases of various string solutions for which the spectroscopy of the anomalous dimensions is elaborated. The string theory side computation of the correlators reproduces the correct anomalous dimensions as predicted by the AdS/CFT correspondence and the exact conformal dependence on the three points at the boundary. Also, a careful analysis has shown that the structure constants of the correlation functions are given by $-g^2\p\Delta_A/\p g^2$. We checked this relation for all considered correlators and found perfect agreement.

The correlation functions we obtained here should be regarded only as a first step. For example, one can consider other string solutions like finite giant magnons. Also, it would be interesting to extend our results to other ``light'' operators, especially the superconformal primary scalar.

%%%%%%%%%%%%%%%%%%%%%%%%%%%%%%%%%%%%%%%%%%%%%%%%%%%%%%%%%%%
\section*{Acknowledgments}
The authors would like to thank P. Bozhilov and H. Dimov for valuable discussions and careful reading of the paper. This work was supported in part by the Austrian Research Funds FWF P22000 and I192, and NSFB DO 02-257.

%%%%%%%%%%%%%%%%%%%%%%%%%%%%%%%%%%%%%%%%%%%%%%%%%%%%%%%%%%%
%%%%%%%%%%%%%%%%%%%%%%%%%%%%%%%%%%%%%%%%%%%%%%%%%%%%%%%%%%%

\end{document}